\documentclass[sigplan,nonacm]{acmart}

\setcopyright{none}
\usepackage[]{hyperref}
\usepackage{mathtools}
\usepackage{braket}
\usepackage{physics}
\usepackage[normalem]{ulem}
\usepackage{bbm}
\usepackage{graphicx}
\usepackage{tikz}
\usepackage{subcaption}
\usepackage{siunitx}
\usepackage{enumitem}
\usepackage{float}
\usepackage{quantikz}

\usepackage{amsmath, amssymb, algorithm,amsfonts, algpseudocode}

\usepackage{graphicx}
\usepackage{textcomp}
\usepackage{xcolor}
\usepackage{booktabs}

\begin{document}

\title{C-Phase-Aware Compilation for Efficient Fault-Tolerant Quantum Execution}

\author{Dhanvi Bharadwaj}
\email{dhanvib@umich.edu}
\orcid{0009-0005-6140-2366}
\affiliation{%
  \institution{University of Michigan}
  \city{Ann Arbor}
  \state{MI}
  \country{USA}
}
\author{Siddharth Dangwal}
\email{siddharthdangwal@uchicago.edu}
\orcid{0000-0002-2483-9105}
\affiliation{%
  \institution{University of Chicago}
  \city{Chicago}
  \state{IL}
  \country{USA}
}

\author{Yuewen Hou}
\email{isaachyw@umich.edu}
\orcid{0009-0004-3035-3197}
\affiliation{%
  \institution{University of Michigan}
  \city{Ann Arbor}
  \state{MI}
  \country{USA}
}
\author{Gokul Subramanian Ravi}
\email{gsravi@umich.edu}
\orcid{0000-0002-2334-2682}
\affiliation{%
  \institution{University of Michigan}
  \city{Ann Arbor}
  \state{MI}
  \country{USA}
}

\begin{abstract}

Achieving practical quantum advantage on fault-tolerant quantum computers (FTQC) is fundamentally constrained by significant spatial and temporal overheads required to execute logical operations on physical hardware. Existing compilers fail to exploit the connection between algorithmic structure and fault-tolerant execution primitives, leaving substantial optimization opportunities unrealized. Additionally, they often rely on coarse-grained, slice-based abstractions that ignore microarchitectural overheads, leading to poor resource utilization.

We present a microarchitecture-aware compilation approach that integrates algorithmic structure directly with lattice surgery (LS) execution. By leveraging the commutativity of C-Phase operations, the method transforms inherently sequential gate sequences into concurrent multi-target interactions, effectively removing artificial dependencies and exposing significant instruction-level parallelism. To enable this, we design a dynamic, event-driven scheduling strategy that accurately models spatial layout and routing constraints, allowing operations to overlap in time while minimizing contention. Compared to the state-of-the-art compiler, our work achieves up to 24.5$\times$~(geomean~3.7$\times$) reduction in execution time, up to 20.8$\times$~(geomean 2.9$\times$) reduction in space-time volume, and up to 2.7$\times$~(geomean 2.4$\times$) reduction in program-failure probability across three regimes of magic-state production. By co-optimizing the circuit reordering and LS primitives, we make executing fault-tolerant programs faster, cheaper, and more reliable on the same hardware.
\end{abstract}

\maketitle

\section{Introduction}
Quantum computing promises to solve problems that are intractable for classical computers. Realizing this potential, however, requires executing circuits with millions of logical operations, far beyond the capabilities of today's noisy intermediate-scale quantum (NISQ) devices~\cite{preskill2018quantum,kim2023evidence}. These systems are limited by decoherence and noise, restricting them to shallow circuits and precluding sustained quantum advantage on practical workloads.

To overcome these limitations, the field is transitioning toward fault-tolerant quantum computing (FTQC)~\cite{shor1996fault,QECIntro,Fowler_2012}, where quantum error correction (QEC) enables reliable execution of deep circuits. This transition introduces a fundamentally new execution model: logical qubits are encoded across many physical qubits, and computation is performed through space-time-intensive logical operations that suppress physical errors~\cite{shor1996fault,beverland2022assessing,Preskill_2025}.

As a result, performance is no longer determined solely by program circuit depth, but by how efficiently logical operations are mapped onto constrained physical resources. This places increasing importance on the compilation and scheduling layers. Despite this, much work in FTQC adopts a decoupled view of the stack, where algorithms are designed and optimized at the program level while compilation is performed independently at the logical execution level.

\begin{figure}[t] 
    \centering
    \includegraphics[width=\columnwidth, trim={0cm 3cm 0cm 3cm}, clip]{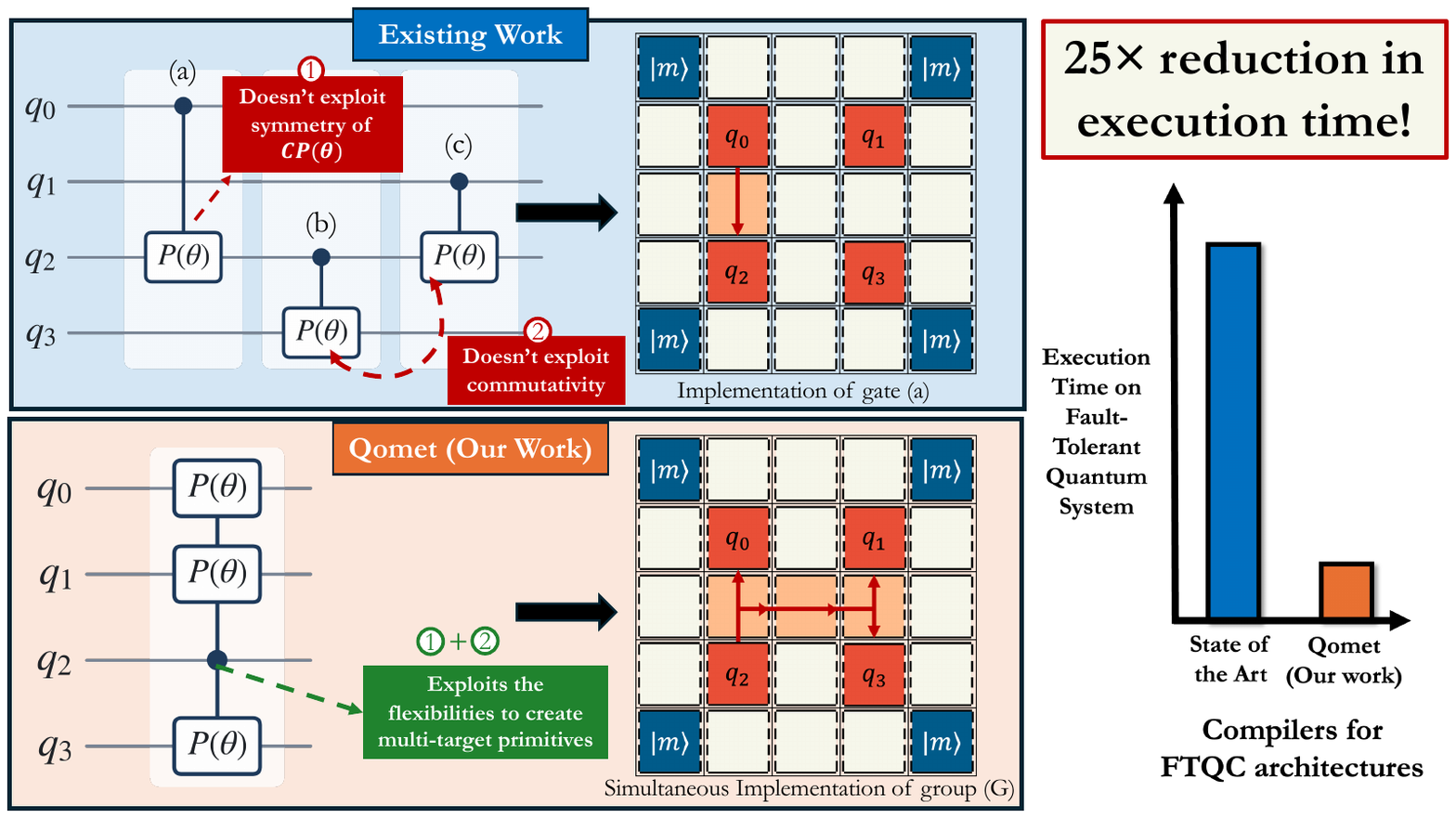}
    \caption{Current compilers impose artificial serialization and layout idling through rigid time-slices. Qomet exploits algorithm-level $C$-phase flexibility and lattice surgery co-design to create multi-target gates, significantly reducing the execution costs.}
    \label{fig:intro_figure}
\end{figure}

\begin{figure*}[t]
    \centering 
    \includegraphics[width=\textwidth, trim={0cm 6.2cm 0cm 3.2cm}, clip]{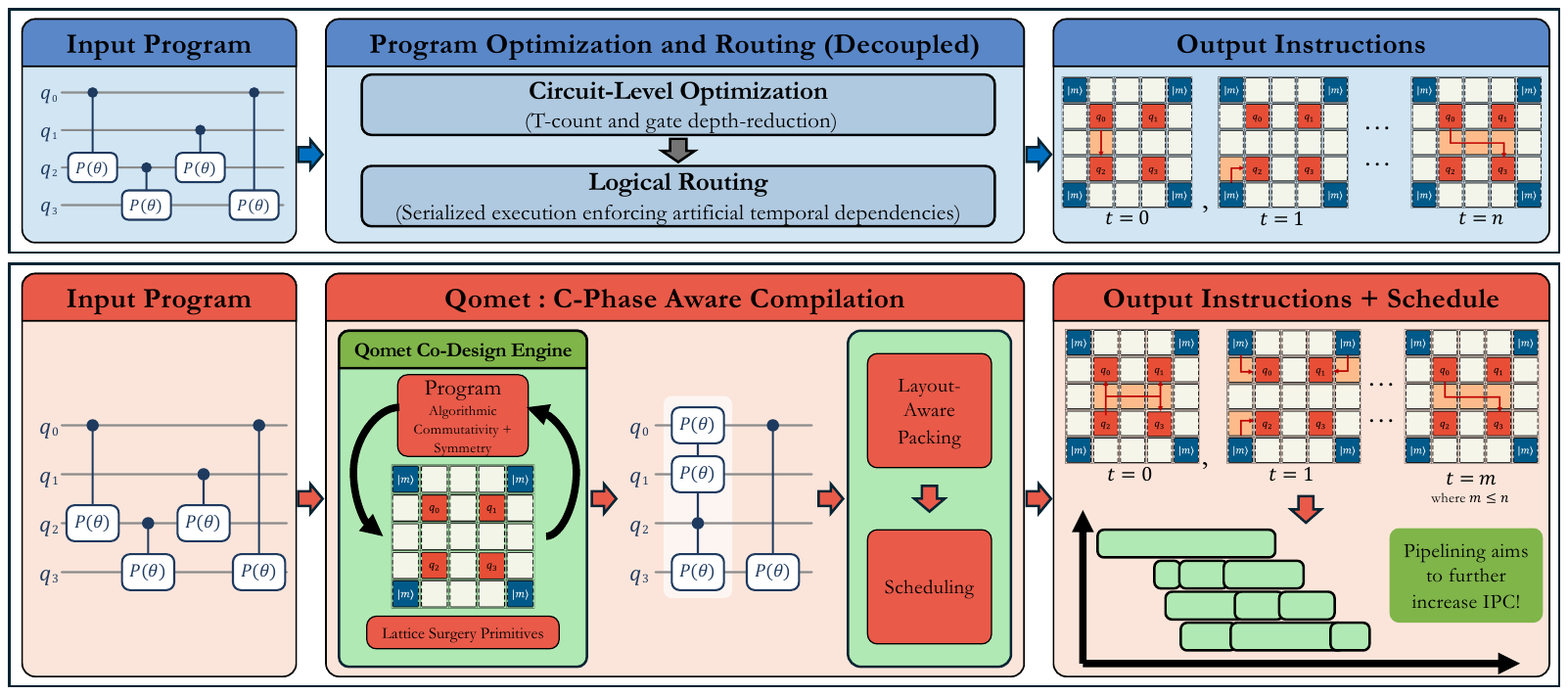}
    \caption{Unlike traditional compilers that decouple optimization from mapping (Top), Qomet (Bottom) introduces a Co-Design Engine that optimizes algorithmic properties with LS primitives. By reinterpreting sequential gate chains as simultaneous multi-target interactions, Qomet bypasses the artificial constraints of slice-based mapping. The resulting pipelined instructions account for realistic microarchitectural overheads (e.g., routing time), enabling high-density instruction execution and a marked reduction in program execution costs compared to state-of-the-art baselines.}
    \label{fig:Qomet_vs_SOTA}
\end{figure*} 

For the surface code, current compilation methods for lattice surgery~\cite{Horsman_2012} (LS) mostly rely on coarse-grained abstractions shaped by architectural models centered around Pauli Product Measurements (PPMs)~\cite{litinski2019game,hua2021autobraid,qulatis}, which are typically tailored for environments without severe quantum and classical resource constraints. Most of these approaches employ a \emph{slice-based} execution model, partitioning circuits into discrete time steps where operations are assumed to execute sequentially over an otherwise idle layout. This abstraction is well-aligned with PPM-based architectures, where operations naturally map to synchronized measurement rounds.

In contrast, this work targets a Clifford+Non-Clifford gate set, which can be more effective when quantum and classical resources are much more limited~\cite{leblond2025quantumresourcecomparisonleading,wang2026taco}. When applied to such gate sets, slice-based execution significantly underutilizes the underlying resources. Additionally, existing cost models prioritize metrics such as T-count or circuit depth, while ignoring routing congestion, resource contention, and the cumulative latency of operations often assumed to be negligible. As fault-tolerant systems scale, these factors increasingly dominate execution time.

We argue that these limitations stem from a failure to fully exploit the interplay between algorithmic structure and fault-tolerant execution primitives. While general-purpose compilers provide broad benefits, prior work has shown that tailoring compilation to a specific program's structure can find substantially larger gains that general-purpose approaches do not see~\cite{jin_structured_2022}. Such program-aware compilation has so far been explored mainly for NISQ devices, and its potential on FTQC is still largely untapped. We observe that key properties of quantum operations, such as the \emph{commutativity} and \emph{symmetry} of controlled-phase (C-Phase) gates, remain largely underutilized in existing compilers (Fig. ~\ref{fig:intro_figure}). Such structure has been exploited productively in the NISQ paradigm, where the objective is reducing two-qubit gate count~\cite{jin_structured_2022}. Under fault tolerance, the relevant cost is lattice-surgery space-time instead, which changes both which transformations pay off and how they must be applied. The opportunity is broad: C-Phase is the core of the Quantum Fourier Transform (QFT)~\cite{campsQuantumFourierTransform2021,baumer2024quantum}, and therefore of phase estimation and Shor's algorithm, while dense C-Phase blocks also form the cost layers of QAOA~\cite{farhi14,bharadwaj2026scalablecliffordbasedclassicalinitialization} and in some instances of Trotterized Hamiltonian Simulation. C-Phase gates are diagonal in the computational basis and therefore commute regardless of qubit overlap, and LS natively supports coordinated \emph{multi-target interactions}, where a single control patch can simultaneously merge with multiple target patches through a shared ancilla region, effectively realizing multiple (commutable) CNOTs in parallel~\cite{litinski2019game,eft_vqa}. Existing compilers are unaware of such macroscopic properties, and often map these interactions into sequential operations, introducing \emph{false dependencies} that inflate runtime. Removing them exposes substantial instruction-level parallelism that conventional compilation strategies leave hidden.

To realize this approach, we present \textbf{Qomet}, a C-Phase-aware compiler for FTQC systems. Qomet integrates algorithm-level transformations with a fine-grained execution model to generate optimized schedules for LS-based architectures (Fig. ~\ref{fig:Qomet_vs_SOTA}). It departs from slice-based abstractions by enabling temporal overlap across operations while managing routing and resource allocation over the 2D lattice. At its core is a realistic cost model that captures spatial constraints, routing latency, and the non-negligible cost of operations such as Clifford gates, allowing the compiler to reason about true execution bottlenecks rather than proxy metrics. Qomet also employs a scheduling framework that minimizes routing conflicts and qubit idling, which is critical for maintaining fidelity in complex programs.

The key contributions of this work are:
\begin{itemize}
    \item \textbf{Qomet Compiler:} A scalable, microarchitecture-aware compiler that integrates algorithmic transformations with realistic execution modeling for FTQC systems.
    
    \item \textbf{Parallelized C-Phase Execution:} A systematic approach to exploit C-Phase commutativity and symmetry to eliminate false dependencies and enable parallel execution.
    
    \item \textbf{Scheduling Framework:} A scheduling strategy that reduces congestion and qubit idling by overlapping instructions across the layout.

    \item \textbf{Performance Gains in Fault-Tolerant Settings:} Qomet delivers consistent and scalable reductions in execution time, achieving up to ${24.5\times}$ speedup over baselines.
\end{itemize}

\noindent
By tightly coupling algorithmic structure with LS-level execution, Qomet unlocks optimization opportunities that are otherwise inaccessible, paving the way toward scalable, high-throughput quantum systems.
\section{Background and Related Work}
\subsection{Quantum error correction and surface codes}

\begin{figure}[t]
    \centering
    \includegraphics[width=\columnwidth, trim={0cm 3cm 0cm 3cm}, clip]{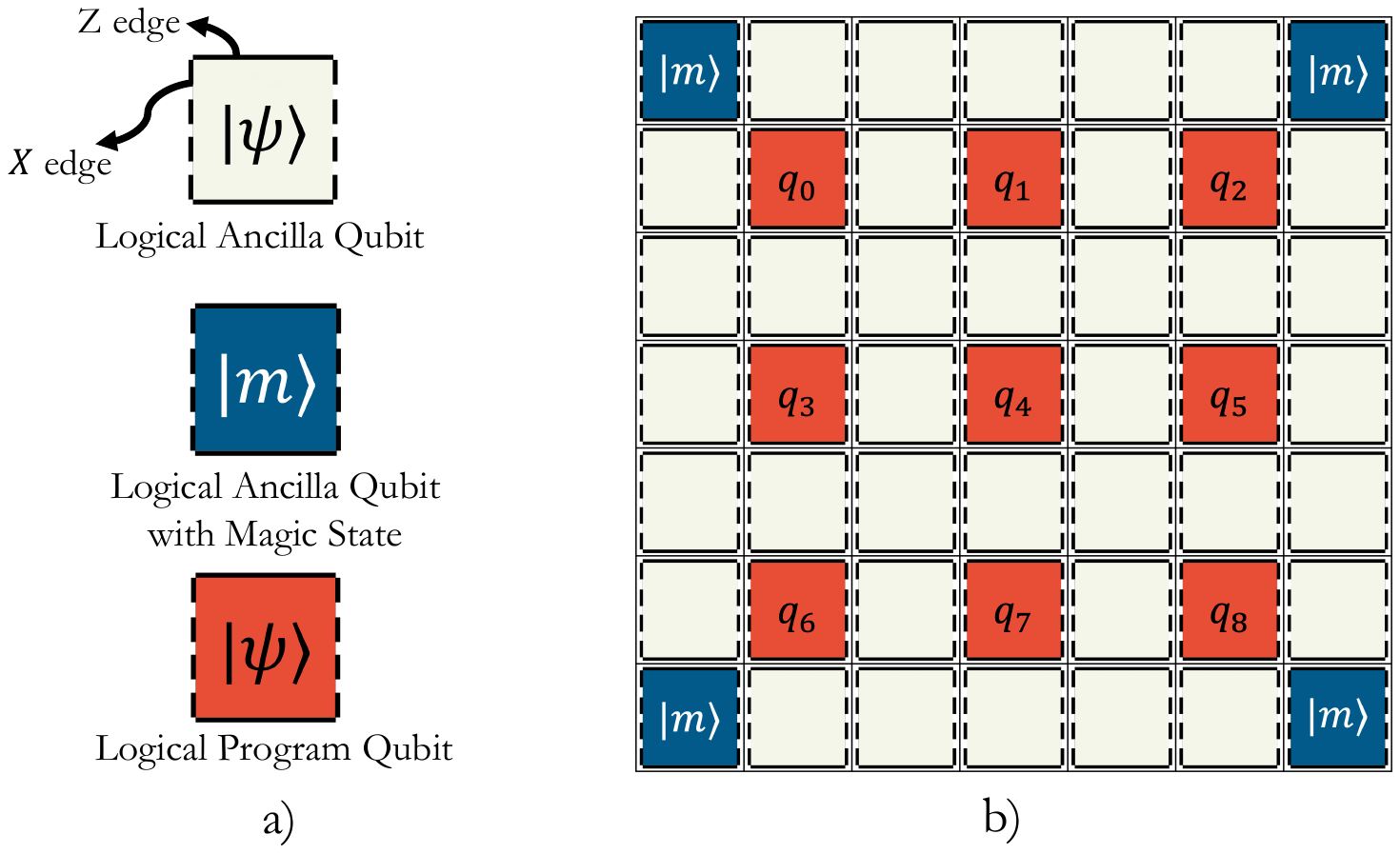}
    \caption{(a) Logical surface code qubits (b) 2-D layout for 9 logical program qubits, where logical operations are realized by coupling program qubit boundaries to intermediary ancilla patches using LS.}
    \label{fig:surface_code_patches}
\end{figure}

Bridging the gap between physical qubit error rates near $10^{-3}$ and the logical error rates near $10^{-15}$ needed for large-scale quantum algorithms requires Quantum Error Correction (QEC)~\cite{lidar2013quantum,Fowler_2012}. The surface code is a leading QEC candidate due to its compatibility with 2D nearest-neighbor architectures, encoding logical qubits as contiguous patches of physical data and measurement qubits~(Fig.~\ref{fig:surface_code_patches}). However, it introduces massive qubit overhead, making architecture a first-order concern. Systems must balance computation, communication, and auxiliary processes like magic state distillation~\cite{litinski2019magic,Beverland_2021} under tight resource constraints. While our work focuses on the surface code, it applies to other two-dimensional topological codes.

\subsection{Lattice Surgery and Patch-Based Layouts}

\begin{figure}[t]
    \centering
    \includegraphics[width=\columnwidth, trim={0cm 5.6cm 0cm 6.4cm}, clip]{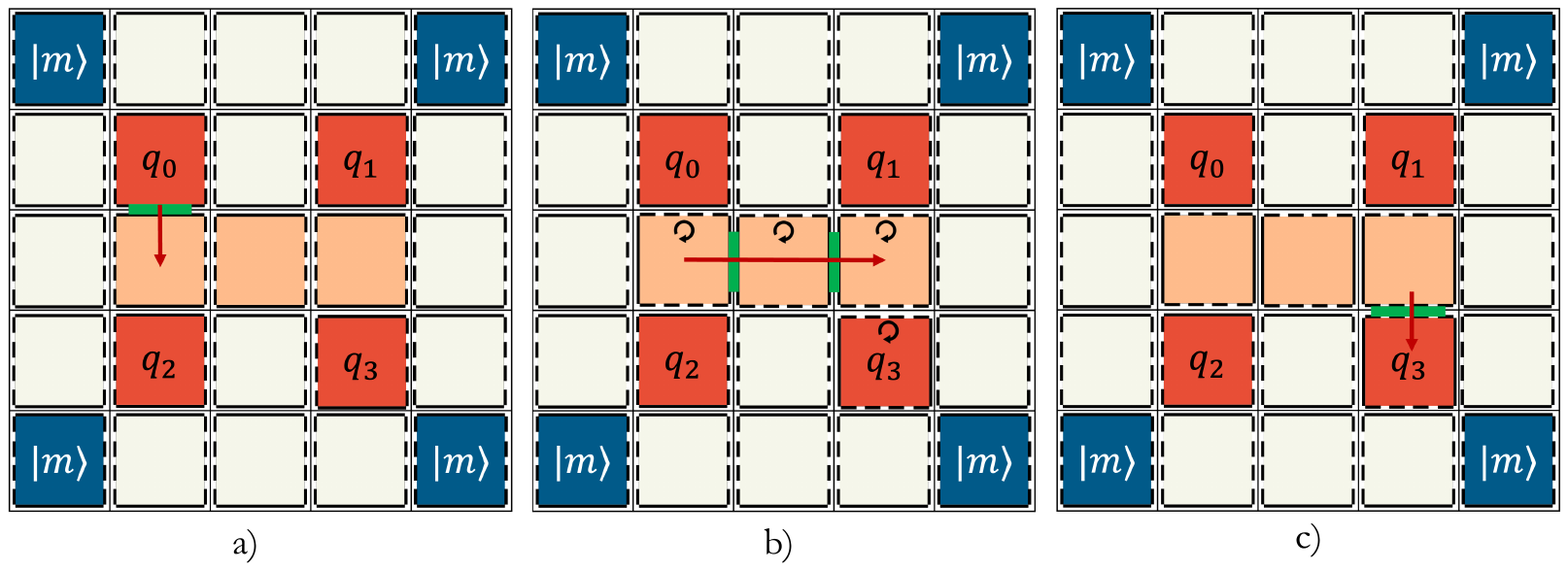}
    \caption{LS instructions for executing Logical CNOT(0,3).
    (a) ZZ Measurement takes 1 clock cycle
    (b) Rotation of each ancilla patch, which can be done simultaneously, takes 1 clock cycle, followed by continuous ZZ measurements, which takes another clock cycle.
    (c) Final XX measurement takes 1 clock cycle. In total, this CNOT takes 4 clock cycles.}
    \label{fig:CNOT_Steps} 
\end{figure}

Lattice Surgery~\cite{Horsman_2012,tan2024sat} (LS) is the leading paradigm for logical operations in surface-code FTQC. Logical qubits are encoded as 2D data patches, and computation proceeds via \emph{merge} and \emph{split} operations that measure joint stabilizers (e.g., $XX$, $ZZ$) across shared patch boundaries. A merge measures the joint operator over $d$ syndrome rounds, and a subsequent split restores the individual patches. Since a merge requires the participating boundaries to be adjacent, and because a compact layout exposes only one $X$ and one $Z$ boundary per patch, operations between distant qubits must be mediated by paths of \emph{ancilla patches}, and mismatched boundary types require patch rotations. As shown in Fig. ~\ref{fig:CNOT_Steps}, a single logical CNOT therefore spans several clock cycles of measurement and rotation. This yields a spatio-temporal tradeoff: sparse layouts ease routing but inflate qubit overhead, while compact layouts save space but cause severe routing congestion. Non-Clifford operations amplify this contention through the cost of producing and routing magic states. Two families of compilation schemes target this instruction set. Pauli-based computation eliminates Clifford gates by commuting them to the end of the circuit and absorbing them into the measurement bases, leaving only $\pi/8$ Pauli product rotations and Pauli product measurements (PPMs)~\cite{litinski2019game}. This removes the time cost of Cliffords entirely, admits a compact layout, and makes resource cost predictable from the logical qubit and $T$ counts alone. What it gives up is parallelism: While PPMs eliminate Clifford gates, they commute them into high-weight multi-qubit rotations that severely restrict parallelism, often reducing throughput to just one gate per layer. This leads to significantly inflated circuit depths, which can drastically increase total execution times~\cite{leblond2025quantumresourcecomparisonleading, wang2026taco}. Additionally, wide multi-patch measurements increase the classical cost, since decoding large patches under correlated logical errors is harder than decoding pairwise merges, and the longer runtime inflates active volume and hence the code distance needed to meet a fixed error budget~\cite{lin2024spatially}.

Direct compilation of Clifford+non-Clifford circuits takes the opposite position, paying for every Clifford explicitly but preserving logical parallelism at the fault-tolerance layer. Neither family dominates, and recent comparisons show the preferred scheme tracks how parallelizable the logical circuit is: nearly serial workloads benefit from Clifford elimination, while dense circuits benefit substantially from direct compilation~\cite{leblond2025quantumresourcecomparisonleading,wang2026taco}. Refinements that schedule compatible PPMs together recover some of this parallelism~\cite{silva2024multiqubit,hirano2025locality}, though the wide support of PPMs limits how far this extends in large registers. We therefore target the direct Clifford+non-Clifford setting, where our central claim is that recovering algorithmic parallelism pushes workloads further into the regime it favors.

Within the Clifford+$T$ model, a lattice-surgery CNOT can be implemented in two ways.
The first method creates a route from a chain of individual square ancilla patches~\cite{eft_vqa} where each step is either a merge between two adjacent patches or a patch rotation that aligns the required boundaries at bends and endpoints. The second method creates the route as a single ancilla region formed by code deformation~\cite{litinski2019game}, which avoids rotations and completes the CNOT in a constant two logical cycles.
However, this lower latency requires irregular patch boundaries, and decoders that support arbitrarily shaped regions, none of which have been fully characterized at the circuit level. We therefore adopt the first model where every step reduces to standard square-patch merges and rotations with well-understood fault-tolerant implementations, giving a conservative, hardware-grounded estimate of execution time that also makes the latency cost of layout geometry explicit.


\subsection{Role of C-Phase Gates}
\label{subsec:cphase}
C-Phase gates are not merely a hardware-specific gate choice but a compact representation of diagonal two-qubit phase interactions. A C-Phase gate is diagonal, \(CP(\theta)=\operatorname{diag}(1,1,1,e^{i\theta})\), and is locally equivalent to a \(ZZ\) rotation up to single-qubit \(R_z\) gates. Since diagonal operators commute regardless of whether their supports overlap, a block of C-Phase gates is fully reorderable and can be treated as a single \emph{phase polynomial}~\cite{Amy_2018} rather than as a fixed sequence. This structure is widespread. It appears directly in QFT and inverse-QFT circuits~\cite{baumer2024quantum}, and hence in QPE and Shor's algorithm~\cite{perlin2026fault}; as edge-wise \(ZZ\) interactions in QAOA cost layers for MaxCut, QUBO, and Ising objectives~\cite{omanakuttan2025threshold}; in Ising and diagonal Hamiltonian simulation, where each Trotter step contains a fully commuting \(ZZ\) block~\cite{ham_sim_ft}; and as the primary mechanism behind QFT-based arithmetic such as adders~\cite{kurt2024scalable}. The rotation angles are arbitrary in these, so the gates are non-Clifford and consume magic-state or rotation-injection resources directly.
A compiler that sees only the decomposed \(\{\mathrm{CNOT},R_z(\theta)\}\) form inherits a rigid chain of two-qubit dependencies, and maps each link onto a separate pairwise merge. A compiler that recognizes the block as a phase polynomial is free to reorder it, and can then exploit the fact that a single control patch may merge with several target patches through one shared ancilla region. The gates are the same; only the second view exposes the parallelism that LS natively supports. Additionally, recently published work supports this motivation by leveraging the phase-polynomial regions that appear frequently in Shor modular exponentiation, QAOA, and Hamiltonian simulation, which constitute more than \(75\%\) of the gates in their representative benchmarks~\cite{phasepoly_isca}.

\subsection{Universal Gate Sets and Rotation Primitives}
\label{subsec:universal_gate_sets}

No QEC code has a transversal universal gate set, so every FTQC architecture supplies non-Clifford operations through an ancilla-mediated protocol. How arbitrary-angle rotations $R_z(\theta)$ are supplied differs sharply across regimes, and that choice determines what the compiler must schedule.

\noindent\textbf{Clifford+$T$.}
\label{subsec:msc_plus_msd}
In the fully fault-tolerant (FFT) regime, each $R_z(\theta)$ is approximated by a Clifford+$T$ sequence. Gridsynth~\cite{gridsynth} requires roughly $3\log_2(1/\delta)$ $T$ gates for target accuracy $\delta$, on the order of hundreds of $T$ gates per rotation at algorithmic precision of $10^{-10}$. Every $T$ gate consumes a magic state. Magic state distillation (MSD)~\cite{bravyi2005universal,litinski2019magic} supplies these magic states from dedicated factories at a spacetime cost exceeding $30d^3$ qubit-cycles per state, against $d^3$ to $3d^3$ for a lattice-surgery Clifford gate. Magic state cultivation (MSC) instead grows a state within a single patch through injection, cultivation, and escape stages, reaching error rates near $2\times10^{-9}$ at $p_{\mathrm{ph}}=10^{-3}$. However, it discards the large majority of attempts and takes roughly several clock cycles per state~\cite{gidney2024magicstatecultivationgrowing}. The compilation consequence is a difference in locality rather than in fidelity: MSD concentrates non-Clifford cost into factories plus the routes that reach them, so parallel $T$ gates demand proportionally more factory area, whereas MSC prepares states in place and lets independent non-Clifford operations proceed concurrently within the ancilla region already reserved for LS.

\noindent\textbf{Clifford+$\phi$ and direct rotation injection.}
\label{subsec:rz_direct_injection}
Early fault-tolerant (EFT) architectures skip synthesis altogether and execute $R_z(\theta)$ directly: a resource state $\ket{m_\theta}=R_z(\theta)\ket{+}$ is prepared non-fault-tolerantly on a single ancilla patch and consumed by gate teleportation, avoiding both synthesis and magic-state factories~\cite{lao2022magic,pqec_rz_injection}. Successive protocols have sharply improved the fidelity of this state. 
STAR reaches $\mathcal{O}(\theta_L^{2(1-\Theta(1/d))} p_{\mathrm{ph}})$ by switching to $T$-gate synthesis for the rare large-angle retries that otherwise dominate the average error~\cite{toshio2026star}. The angle dependence is what matters here: since Trotterized and phase-estimation circuits have many small-angle rotations, these scalings make direct injection progressively more favorable exactly where such rotations are most numerous.
Two properties of this regime impact scheduling. First, teleportation succeeds only half the time, and on failure the rotation must be retried with the angle doubled, so the latency of an individual rotation is a distribution rather than a fixed cost. Second, since the preparation is local, independent rotations can be issued concurrently within the ancilla region already reserved for LS~\cite{hirano2025locality}. Recent architectures combine both regimes under a Clifford+$T+\phi$ gate set, choosing per gate by rotation angle~\cite{toshio2026star}.
The costs used in our experiments are detailed in~\autoref{subsec:fine_grained_cost_model_numbers}.

\subsection{Prior Compilation Work}
\label{subsec:prior_work}
Logical compilers for surface-code architectures typically address the NP-complete Surface Code Mapping and Routing (SCMR) problem by optimizing qubit placement and ancilla routing~\cite{scmr}. State-of-the-art Clifford+$T$ compilers predominantly rely on slice-based abstractions, partitioning execution into discrete, synchronized time steps~\cite{dascot,LSC_Compiler}. In practice, LS operations exhibit heterogeneous latencies (CNOT, $T$-injection, patch rotations), but slice-based models assume uniform slice duration, so each step is bottlenecked by its slowest operation. More fundamentally, enforcing global slice completion before advancing blocks execution even when resources are free, serializing otherwise independent operations. These models also treat Clifford-level operations such as $H$, $S$, and grid resets as negligible, though each incurs non-trivial latency at distance $d$. Separately, prior compilers rarely treat algorithmic structure as an optimization goal~\cite{harkness2026ftcircuitbench,chamberland2026patchwork}. Structure-aware transformations are well established in NISQ compilation~\cite{jin_structured_2022}, but target two-qubit gate count rather than lattice-surgery space-time, and LS's native support for multi-target merges is left unexploited. Qomet addresses both gaps with a cycle-accurate, dependency-aware execution model that treats the lattice as a continuous spatio-temporal resource.

\section{Multi-Target C-Phase Execution and Cycle-Accurate Scheduling}

Our key insight is that multi-target execution is a native LS primitive that lets one control patch merge with several targets at once, removing the false dependencies that gate-level compilation imposes on commuting C-Phase chains to make program execution faster. Our secondary insight is that cycle-accurate scheduling amplifies this gain by modeling routing and protocol costs at cycle granularity, letting operations that slice-based schedulers serialize overlap in time.


\begin{figure}[t]
  \centering
  {\scriptsize
  \begin{quantikz}[row sep={0.55cm,between origins}, column sep=0.12cm,
                   every node/.append style={inner sep=1pt}]
    & \ctrl{1} & \midstick[2,brackets=none]{$=$}
    & \ctrl{1} & \qw & \ctrl{1} & \gate{R_Z(\theta/2)} & \\
    & \gate{P(\theta)} &
    & \targ{} & \gate{R_Z(-\theta/2)} & \targ{} & \gate{R_Z(\theta/2)} &
  \end{quantikz}}
  \caption{Decomposition of $\mathrm{CP}(\theta)$ into two CNOTs and three $R_Z$ rotations.}
  \label{fig:CP_Decomposition}
\end{figure}

\subsection{Multi-target execution to eliminate false dependencies}
\label{subsec:multi_target_execution}

\subsubsection{False Dependencies in C-Phase Chains}
\begin{figure}[t]
    \centering
    \includegraphics[width=\columnwidth, trim={3cm 3cm 2.5cm 3cm}, clip]{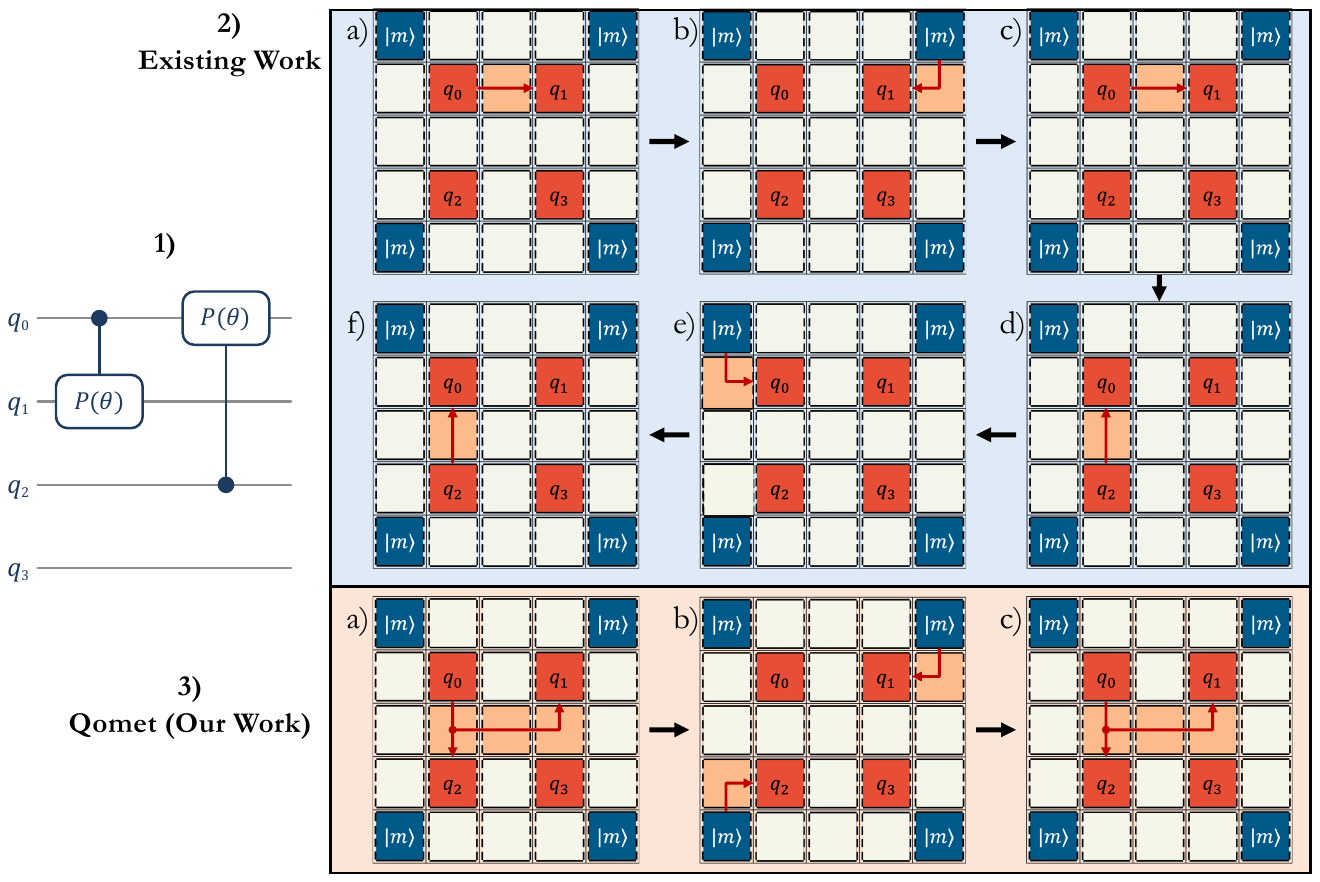}
    \caption{Eliminating false dependencies via CP-aware compilation. 
   (1) Example QAOA circuit with two cost-layer two-qubit phase gates. These ZZ rotations decompose exactly into CNOT--$R_z(\theta)$--CNOT \cite{jin_structured_2022}.
    (2) Existing approach: despite spatial independence, the compiler enforces serialization. Slices (2a–c) show the CNOT–$R_z(\theta)$–CNOT decomposition of $\mathrm{CP}(0,1)$, while (2d–f) show $\mathrm{CP}(2,0)$ under the same constraints. Although operations such as (2d) and (2f) could overlap with (2a) and (2c), false dependencies force sequential execution.
    (3) Qomet: exposes parallelism via multi-target C-Phase execution. Slices (3a) and (3c) implement $\mathrm{CNOT}(0,1)$ and $\mathrm{CNOT}(0,2)$ with one route, while (3b) implements the magic states on both the targets, in the same slice.
    {Each slice denotes the lattice-surgery operations executed concurrently. Every operation is a logical CNOT implemented according to Fig.~\ref{fig:CNOT_Steps}, and code-cycle durations are modeled by Qomet's cost model (Sec.~\ref{sec:cost_model})}.}
    \label{fig:false_dependency}
\end{figure}

C-Phase is decomposed into a $CNOT$-$R_z(\theta)$-$CNOT$ block, plus local corrections (Fig. ~\ref{fig:CP_Decomposition}). Traditional compilers serialize chains such as $\{CP(C, T_i)\}_{i=1}^n$, requiring each C-Phase to complete before the next begins, even when gates can be executed in the same circuit layer and routing space is available. In Fig.~\ref{fig:false_dependency}, a compiler will serialize $CP(0,1)$ and
$CP(2,0)$ because they share qubit $q_0$. We observe this is a \emph{false dependency} created by existing compilers which optimize the circuit and routing separately. Taking $q_0$ as the control of both lets their CNOTs execute as one multi-target CNOT,
so the two target $R_z(\theta)$ rotations, on different qubits, also run in parallel.

\subsubsection{Implementation of Multi-Target Operations}

\begin{figure}[t]
    \centering
    \includegraphics[width=0.8\columnwidth,clip]{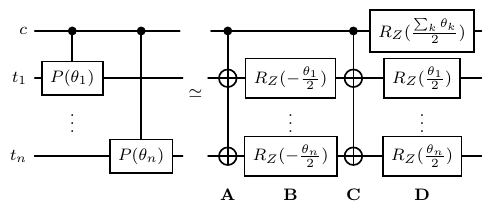}
    \caption{Multiple C-Phase gates sharing control $c$ (left), regrouped into a
  multi-target block (right). Since the $R_Z$ corrections are diagonal, they
  commute past the $ZZ$ blocks, so the corrections on the control combine
  into a single $R_Z\big(\sum_k\theta_k/2\big)$.}
    \label{fig:4-stage-CPhase}
\end{figure}

A C-Phase gate is a $ZZ$ rotation (two CNOTs around a target $R_z$)
followed by local $R_z$ corrections (Fig.~\ref{fig:CP_Decomposition}). With
operators applied right to left,
\begin{equation}
CP_{c,t}(\theta) = e^{i\theta/4}\,
\underbrace{R_z^{c}\big(\tfrac{\theta}{2}\big)R_z^{t}\big(\tfrac{\theta}{2}\big)}_{\text{corrections}}\,
\underbrace{\mathrm{CX}_{c\to t}\,R_z^{t}\big(\!-\tfrac{\theta}{2}\big)\,\mathrm{CX}_{c\to t}}_{ZZ\text{ rotation}},
\label{eq:cp_decomp}
\end{equation}
where
\begin{equation}
R_z(\phi)=e^{-i\phi Z/2}.
\label{eq:rz}
\end{equation}
When multiple C-Phase gates share a control $c$, their CNOTs all commute, so they
merge into one multi-target CNOT
\begin{equation}
F=\prod_k \mathrm{CX}_{c\to t_k}.
\label{eq:multi_cnot}
\end{equation}
Since
\begin{equation}
F\,Z_{t_k}\,F=Z_c\,Z_{t_k},
\label{eq:f_conj}
\end{equation}
placing all target rotations between two copies of $F$
can apply every $ZZ$ rotation at once:
\begin{equation}
\prod_k CP_{c,t_k}(\theta_k)=e^{i\Theta/4}\,D\,C\,B\,A,
\qquad \Theta=\sum_k\theta_k.
\label{eq:multi_cp}
\end{equation}
We define these stages in Fig.~\ref{fig:4-stage-CPhase} that execute in order:
(A)~multi-target CNOT $F$;
(B)~$R_z^{t_k}(-\theta_k/2)$ on all targets in parallel;
(C)~$F$ again; and
(D)~$R_z^{t_k}(\theta_k/2)$ on each target plus a \emph{single}
$R_z^{c}(\Theta/2)$ on the control.
Stages A--C realize the $ZZ$ rotations, and Stage D completes them into C-Phase
gates. Merging the $n$ corrections on the control qubit in Fig.~\ref{fig:CP_Decomposition} into one rotation saves $n-1$
non-Clifford operations. $R_{ZZ}$ blocks such as QAOA's cost layer need only
Stages A--C, with angle $\phi_k$ in Stage B, since
\begin{equation}
e^{-i\phi Z_cZ_t/2}=\mathrm{CX}_{c\to t}\,R_z^t(\phi)\,\mathrm{CX}_{c\to t}.
\label{eq:rzz}
\end{equation}

\subsection{Fine-Grained Cost Modeling and Scheduling}
We replace existing logical time-slice abstractions with a cycle-accurate model.
LS primitives have different costs that existing works don't account for: a merge occupies $d$ syndrome rounds, $H$,
$S$, and patch resets take $O(1)$ cycles, and rotation latency depends on
the injection or distillation protocol. Slice-based compilers make every
operation in a slice wait for the slowest one. Our scheduler instead
tracks the layout at cycle granularity and frees each patch when its
operation completes, so dependent instructions start immediately; only
multi-target groups are issued as atomic units, so that their planned
routes stay available (~\autoref{sec:pipelining}). The same model serves
every regime by treating magic states as architectural resources rather
than fixed gate costs. STAR rotations are cheaper and random
under RUS; MSD requires long-range routing from
factories; and MSC grows each state near its target, trading routing for
cultivation latency that the scheduler overlaps with other instructions. 

\section{Design of Qomet}
\label{sec:features}

We detail the design of Qomet in this section. Our fine-grained cost model (\autoref{sec:cost_model}) drives three optimization passes: \emph{multi-target grouping} (\autoref{sec:fan_out_formation}), which identifies parallelizable interactions; \emph{routing-aware packing} (\autoref{sec:packing}), which maximizes utilization of available ancilla resources; and \emph{$R_z(\theta)$ implementation} (\autoref{sec:rz_realization}). We then detail our pipelined scheduler (\autoref{sec:scheduler}) that removes slice barriers with cycle costs per instruction. 


\subsection{Fine-Grained Latency Cost Modeling}
\label{sec:cost_model}

Existing compilers assign uniform costs to LS operations, abstracting away LS-level effects \cite{dascot,hofmeyr2025puremagic,LSC_Compiler}. Qomet instead precisely models routing, patch orientation, and the coordination overheads incurred during compilation. We model cost over a route represented as a path \(P=(v_0,v_1,\ldots,v_k)\) on the layout, where each patch carries an orientation. Traversing the path, consecutive moves in the same direction are grouped into straight segments. For each segment, we add one clock cycle for movement; if any cell in the segment is not already in the required orientation, we rotate the entire segment once and add one additional cycle. The final LS merge costs one cycle. Adding all the movement, rotation, and merge costs yields the total cycle count for the route, and the resulting cell orientations are committed for subsequent routes. Costs therefore depend on both the selected route and the state of the layout at the time of routing.


\subsection{Multi-Target Group Formation}
\label{sec:fan_out_formation}

This section details the process of creating groups (\(\{G_1, \ldots, G_K\}\)) of single-control-multi-target C-Phase gates. Each group \(G_k\) has a single control qubit \(c_k\) whose patch simultaneously merges with all target patches \(T_k = \{t_j^{(k)}\}\). A group's execution time depends on two costs. The first is the routing time \(L(u,v)\) of a gate, computed as the cycle-minimizing shortest route through the ancilla patches between \(\phi(u)\) and \(\phi(v)\) under our cost model (\autoref{sec:cost_model}), for qubit layout \(\phi: q \mapsto (r, c)\). The second is the cost \(R(q)\) of applying the \(R_z(\theta)\) on qubits. This cost depends on where qubit \(q\) sits and on which protocol creates these non-Clifford gates. Under MSD, it is the route from the nearest reachable factory, paid once per \(T\) gate. Under MSC and STAR, it is the probabilistic cost, \(q\), and STAR splits this cost across all available ancilla patches~\cite{eft}. Stages A and C are each bounded by their slowest CNOT and Stages B and D by their slowest target, so a group with control \(c\) and targets \(T\) takes
$\tau(c,T) = 2\max_{t\in T} L(c,t) + \max_{t\in T} R(t) + \delta_D \max_{q\in\{c\}\cup T} R(q)$
cycles, where $\delta_D = 1$ for exact C-Phase blocks (e.g., QFT), which also
run Stage~D, and $\delta_D = 0$ for RZZ blocks such as QAOA's cost layer. The
objective is therefore to form groups that spend the fewest cycles per gate.

At each step, every qubit $q$ with remaining gates is a candidate
control. Its remaining gates form one group, with its interaction partners
$N(q)$ (qubits sharing a remaining gate with $q$) as targets. Each candidate
is scored by its cost per gate, i.e., the group's latency divided by the
number of gates it executes. This favors large groups
unless a costly target, through a long route or slow rotation, raises the
latency of the whole group. The lowest-scoring qubit becomes the control
$c_k$ and takes all its remaining gates. Since
only target qubits need Stage~B, this score selects qubits with expensive
rotations (e.g., far from a factory or with few free neighbouring patches)
as the control, where their rotation is skipped (RZZ) or merged into a
single Stage~D rotation. Since the diagonal phase block is symmetric (\autoref{subsec:cphase}), either endpoint can be the control, which makes this a scheduling choice. Algorithm~\ref{alg:fan_out_formation} also enforces that no qubit appears in multiple roles within a group, so the multi-target merge is physically valid.

\begin{algorithm}[t]
    \caption{Multi-Target Group Formation}
    \label{alg:fan_out_formation}
    \begin{algorithmic}[1]
    \Require Gate list $\mathcal{G}$, qubit map $\phi$, grid $\mathcal{L}$
    \Ensure Ordered multi-target groups $[G_1, \ldots, G_K]$
    \For{each gate $g = (u,v) \in \mathcal{G}$}
        \State $L(u,v) \gets \textsc{PathCost}\bigl(\textsc{ShortestPath}(\phi(u), \phi(v), \mathcal{L})\bigr)$
        \State $R(u) \gets \textsc{RzCost}(u, \phi, \mathcal{L})$;\quad $R(v) \gets \textsc{RzCost}(v, \phi, \mathcal{L})$
    \EndFor
    \While{$\mathcal{G} \neq \emptyset$}
        \For{each qubit $q$ appearing in $\mathcal{G}$}
            \State $N(q) \gets \{ v : (q,v) \in \mathcal{G} \text{ or } (v,q) \in \mathcal{G} \}$
            \State $s(q) \gets \tau\bigl(q, N(q)\bigr) / |N(q)|$
        \EndFor
        \State $c_k \gets \arg\min_q s(q)$, breaking ties by larger $|N(q)|$
        \State $G_k \gets \{ (c_k, t) : t \in N(c_k) \}$, ordered by decreasing $R(t)$
        \State $\mathcal{G} \gets \mathcal{G} \setminus \{ g \in \mathcal{G} : c_k \in g \}$
    \EndWhile
    \end{algorithmic}
\end{algorithm}

\subsection{Heuristic Packing via Steiner Trees and BFS}
\label{sec:packing}
After grouping, many groups have space that is underutilized. Qomet therefore \textit{packs} compatible gates from later groups whenever unused ancilla resources permit.
\textbf{Phase 1: Steiner tree for the primary interaction.}
For each group $G_k$ with control $c_k$ and targets $\{t_i\}$, Qomet creates an approximate Steiner tree over a pruned graph containing only the relevant data and ancilla patches. The tree defines the reserved routing footprint $\mathcal{B}_k$. Since all merges within $\mathcal{B}_k$, aim to execute in parallel, group latency is the longest root-to-target path in the tree rather than the sum of individual paths. After the tree is built, each route is costed under our
model (including any rotation of shared cells), and targets that
cannot be routed jointly fall back to point-to-point routing.
\textbf{Phase 2: BFS-based gate packing.}
Qomet greedily packs later gates into $G_k$ by computing BFS shortest paths within the remaining ancilla space of $\mathcal{B}_k$. Each committed gate updates the footprint until no further packing is possible, and the final group latency is $\max_{g \in G_k} L(g)$.

\subsection{Implementing $R_z(\theta)$}
\label{sec:rz_realization}

Qomet treats $R_z(\theta)$ implementation as a first-class scheduling constraint and supports both regimes, adapting routing strategy accordingly.

\subsubsection{STAR (EFT)}
\label{sec:rz_injection}
Under Clifford+$\phi$, each target receives its $R_z(\theta)$ via continuous-angle injection~\cite{toshio2026star}: an ancilla patch near the data qubit locally prepares the resource state and teleports it with a logical CNOT (2 cycles). This is architecturally efficient, since the injection ancilla sits near the data qubit, which allows for short routes. All targets in a C-Phase group inject simultaneously during stage B, each using neighboring ancillas. The preparation costs for this are derived in \autoref{subsec:fine_grained_cost_model_numbers}. 
Recent improvements in this technique utilize RUS protocols to generate arbitrary magic states $|m_{\theta}\rangle$~\cite{toshio2026star,pqec_rz_injection,toshio2025practical}. These facilitate direct $R_z(\theta)$ injection, eliminating the need for costly decomposition.

\subsubsection{$T$ via Magic State Distillation (FFT)}
\label{sec:t_routes}
Under Clifford+$T$, each $R_z(\theta)$ is first synthesized into a \{T, S, H\} sequence via GridSynth~\cite{gridsynth} at a precision \(p\). Every $\ket{T}$ must then be routed from an ancilla patch holding a prepared magic state to the target qubit, evaluated under the same cost model. Accounting for the full decomposition, one \(R_z(\theta)\) costs \(\tau(t_j)=\tau_{\mathrm{route}}N_T + T_S N_S + T_H N_H\), where \(\tau_{\mathrm{route}}\) is the routing cost, \(T_S\) and \(T_H\) are the expected latencies of \(S\) and \(H\) gates, and \(N_T\), \(N_S\), and \(N_H\) are their respective counts. \textbf{Cost-aware magic-state selection.}
Assigning each target to its nearest magic-state patch by Manhattan distance is a common heuristic but can be suboptimal: the closest patch may connect via a winding, high-rotation-cost path while a slightly farther one offers a straight corridor at lower cost. Qomet sorts patches by Manhattan distance as a cheap pre-filter, evaluates the full routing cost for the top-\(k\) candidates, and assigns the patch minimizing \(\tau_{\mathrm{route}}\). 


\subsubsection{$T$ via Magic State Cultivation (FFT)}
MSC grows a high-fidelity magic state in one ancilla patch next to the target (\autoref{subsec:msc_plus_msd}) and consumes it by gate teleportation. Like injection, it needs no factory or long-range routing. Cultivation is a RUS protocol, so each $\ket{T}$ state has a sampled latency (\autoref{subsec:fine_grained_cost_model_numbers}).

\subsection{Pipeline-Scheduling in Qomet}
\label{sec:scheduler}

In this section we detail how Qomet schedules LS instructions efficiently on the layout, managing routing and resource allocation to execute a program. Qomet places every instruction of the program on a single timeline - the
$CP(\theta)$ groups built in \autoref{sec:packing} as well as the surrounding
non-$CP(\theta)$ instructions (CNOTs, $R_z(\theta)$s and single-qubit
Cliffords).

\subsubsection{Event-Driven Cycle-Accurate Execution}
\label{sec:event_driven}

We model execution as a discrete-event simulation on the layout. Each
operation becomes an event with a precise duration, and a priority queue
advances time to the next completion, releasing patches and dispatching
the next instructions. Cycle counts are computed under the cost model of
\autoref{sec:cost_model}, calculated on the state of the layout with
an orientation tracker that is updated after each route. Routes for CNOTs and
magic-state deliveries are computed when the operation is dispatched,
against the live state of the layout.

\subsubsection{Scheduling $CP(\theta)$ Groups}
\label{sec:pipelining}
Once Qomet creates routable groups of $CP(\theta)$s and packs them into
slices, it schedules each slice as a unit, so that the consecutive routes
planned for a group are never broken. A slice consists of one multi-target
group, routed along its Steiner tree, and the point-to-point $CP(\theta)$s
packed alongside it. All of Stage~A
starts at the same cycle. The group performs its multi-target CNOT
along the Steiner tree through the ancillas, and each packed gate performs
its CNOT on its own path. Once Stage~A is fully complete, the scheduler issues that
target's Stage~B $R_z(\theta)$ on the layout. For the group, the multi-target
CNOT completes as one operation, so all of its targets start Stage~B
together. A packed gate's Stage~B starts as soon as its own route is done.
Only when all of the group's Stage~B $R_z(\theta)$s are done do we perform
Stage~C, reusing the same tree cells as Stage~A. Stage~C waits for all of
Stage~B so that the CNOTs from the shared control start at the same cycle
and are again executed as one multi-target CNOT. A packed gate instead
proceeds to Stage~C as soon as its own Stage~B completes, reusing its
Stage~A path. Stage~D then proceeds on each target as soon as its Stage~C
CNOT is done, just like Stage~B. We also note that the control's
$R_z(\tfrac12\sum_k\theta_k)$ is issued once every Stage~C CNOT sharing
that control has completed.


\subsubsection{Pipelining Non-$CP(\theta)$ Instructions}
\label{sec:interleaving}

Instructions outside $CP(\theta)$ blocks are dispatched as soon as possible,
respecting program order where an instruction waits for the
previous operation on each of its qubits. A CNOT is routed on the
layout when it is dispatched, $R_z(\theta)$ is scheduled exactly as a
Stage~B /D rotation (\autoref{sec:rotation_scheduling}), and $H$ and $S$ follow space and time costs detailed in \autoref{subsec:fine_grained_cost_model_numbers}.  


\subsubsection{Scheduling the Rotation Stage}
\label{sec:rotation_scheduling}

Our scheduler adapts to different magic-state protocols. For STAR, Stages~B and D require only free ancilla patches adjacent
to the program qubit. The scheduler reserves at most four adjacent patches that are free
when the rotation starts, splits the sampled injection cost
$\tau_{R_z(\theta)}$ across them, and holds them until the rotation
completes~\cite{eft_vqa}. This makes implementing an $R_z(\theta)$ fast, with negligible routing
overhead. For MSC, the scheduler instead allocates a single free adjacent
patch, on which $\ket{T}$ is cultivated and then teleported into the
program qubit. In MSD, each $R_z(\theta)$ is synthesized into a Clifford+$T$
sequence, and each $\ket{T}$ is routed from a factory. The scheduler
considers $k$ factories closest to the target and takes the shortest-cycle route
that is free on the current layout ($k=8$ by default). When several Stage~B and D operations
contend for the same factories, a rotation that cannot get its
neighbouring patch or a free route waits for the next event and is retried
on the updated layout.  This gives a contention-aware $\ket{T}$-state
delivery mechanism. 


\section{Methodology}
\subsection{Baseline Compilation Frameworks}
\label{subsec:baseline_frameworks}
We compare Qomet against two baselines spanning different configurations and fault-tolerant architectures. 

\textbf{DASCOT.} Our primary state-of-the-art baseline is DASCOT, a dependency-aware compiler for MSD architectures that formulates SCMR with the objective of minimizing the total number of execution slices, via simulated annealing \cite{dascot}. We use wisq v0.2.4, unmodified, in mapping-and-routing-only mode (\texttt{--mode scmr}). The mapping-and-routing budget (\texttt{-tmr}$=24$ hrs) is split
equally between the two phases. A run that exceeds this time
produces no schedule and is reported as a timeout. DASCOT targets only MSD, but to also evaluate it under STAR and MSC, we route a CNOT-only version of each circuit with DASCOT and price every $R_z(\theta)$, $T$, and single-qubit Cliffords with the same sampled MSC or injection cost Qomet uses, so differences reflect compilation rather than magic-state cost assumptions.


\textbf{PureMagic.} DASCOT's search-based routing is slow and scales poorly. We therefore add PureMagic~\cite{hofmeyr2025puremagic} as a second, more scalable baseline. PureMagic is a dynamic LS scheduler designed for MSC in the PPM model. Instead of optimizing gate orderings, it uses greedy Steiner-forest packing fast enough to make scheduling decisions within a logical cycle. To evaluate it on all three architectures, we adapt it to run with instantly available magic states and price its schedule with the same MSD, MSC, and STAR cost models used for Qomet and DASCOT. PureMagic absorbs Clifford gates into the T gates' Pauli rotations up to a maximum Pauli-product weight $\omega$. As done in their work, we use $\omega=1$ (all Pauli products stay single-qubit) for the Clifford+T execution model and $\omega=\infty$ (all Cliffords absorbed) for PBC.

\subsection{Evaluation Metrics}
    \noindent\textbf{Reduction in Execution Time.}
    This metric measures the ratio of the total execution time of a baseline's compiled program to Qomet's, where execution time is the number of logical clock cycles needed to run the full program. A value $>1.00$ indicates Qomet executes a program faster than the baseline.
    
    \noindent\textbf{Reduction in Space-Time Volume.}
    Space-time volume is the standard lattice-surgery cost metric, capturing the qubit–runtime trade-off that dominates fault-tolerant resource requirements~\cite{litinski2019game,beverland2022assessing,litinski2022activevolumearchitectureefficient}. We define this metric as the ratio of a baseline compiler's space-time volume to Qomet's. Space-time volume is $V = n \cdot T + \sum_{\text{ops}} c_{\text{op}} \cdot t_{\text{op}}$, where $n$ is the number of logical program qubits, which perform syndrome extraction every cycle to protect the logical state, plus each operation's non-data patches $c_{\text{op}}$ (routing and magic-state patches) times its duration $t_{\text{op}}$.
  
    \noindent\textbf{Reduction in Program Failure Probability}
    Better compilation should also directly improve program reliability, since every patch
    accumulates logical errors for every cycle it remains active, including
    while idling~\cite{Fowler_2012, acharyaQuantumErrorCorrection2025}. Following
    standard surface-code resource estimation work~\cite{Fowler_2012,
    litinski2019game, gidney2021factor, beverland2022assessing}, we assume
    each patch fails independently with logical error rate $p_L$ per clock
    cycle. For a program with space-time volume
    $V$, the probability of at
    least one logical fault during execution is
    $P_{\text{fail}} = 1 - (1 - p_L)^{V}$. We make two assumptions:
    (i)~Both compilers run the same logical program with the same gadget
    library (LS merges and splits, patch rotations, $R_z(\theta)$ injection,
    and MS teleportation), code distance, and cost model.
    (ii)~All patches have the same per-cycle logical error rate $p_L$, so the
    compilers differ only in scheduling, not gadget fidelity. Resource-state preparation errors
    are common to both and excluded. We report
    $P_{\text{rel}} = P_{\text{fail}}^{\text{Baseline}} / P_{\text{fail}}^{\text{Qomet}}$,
    where $P_{\text{rel}} > 1$ favors Qomet.

\subsection{Benchmark Applications}
Our suite spans structured and irregular circuit behavior across regimes. 
\textbf{QFT.} A highly structured but dependency-heavy workload spanning a broad range of rotation precisions, and a core subroutine of many algorithms including Shor's factoring. Its long-range interaction pattern and heterogeneous gate latencies stress cycle-accurate routing and execution overlap. \textbf{QAOA.} A variational optimization workload whose cost layers consist of commuting $ZZ$ rotations with strong symmetry structure, making it well suited to evaluating multi-target parallelization and layout-aware grouping. QAOA is increasingly treated as an EFT candidate, since useful problems are expected to require error correction~\cite{perlin2026fault,omanakuttan2025threshold}. We evaluate depth-$p=2$ QAOA over 10 Max-Cut instances generated from Erd\H{o}s--R\'enyi graphs with edge probability 0.5.  \textbf{Simulation and arithmetic workloads.} To test workloads with limited or no C-Phase structure, we evaluate on jellium in the plane-wave dual basis~\cite{babbush2018low,mcclean2020openfermion},
the Sherrington--Kirkpatrick model~\cite{sherrington1975solvable}, and circuits from prior
work~\cite{phasepoly_isca}: adders~\cite{kurt2024scalable}, modular
arithmetic~\cite{ruiz2017quantum}, multi-controlled-gate decompositions, and
transverse-field Ising simulation~\cite{ham_sim_ft}.

\subsection{Layout and Magic-State Configurations}
\begin{figure}[t]
    \centering
    \includegraphics[width=\columnwidth, trim={0cm 5.8cm 0cm 5.4cm}, clip]{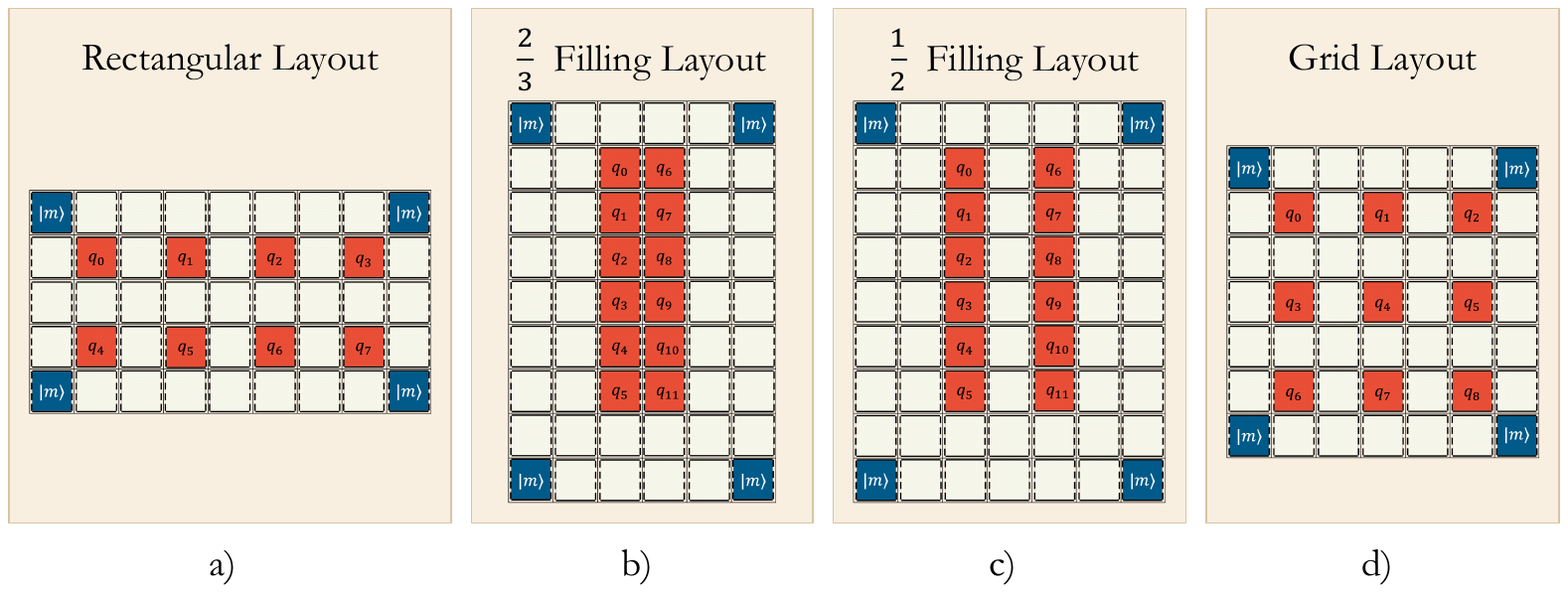}
    \caption{Layouts proposed in prior work with the MS-starved density: (a) Rectangular (adds a perimeter of ancilla patches to the layout in \cite{dascot} to provide dedicated area for MS patches) (b) 2/3-filling floorplan~\cite{kobori2025lsqca},(c) 1/2-filling~\cite{kobori2025lsqca}, and (d) Grid~\cite{dascot}.}
    \label{fig:selected_layouts}
\end{figure}

We evaluate four layouts that differ in density and routing space (\autoref{fig:selected_layouts}). Their asymptotic program-qubit densities are $1/5$, $1/3$, $2/7$, and $1/4$, covering both relaxed and compact layouts. We also test two densities of available magic states. In \emph{MS-abundant}, available magic states scale with program size along the boundary, modeling ample supply. In \emph{MS-Starved}, four corner patches are assigned regardless of size, which stresses routing under MS scarcity. Following~\cite{dascot}, factories sit in a separate region whose output rate we do not model, which also does not add to the space-time volume. 


\subsection{Fine-Grained Clock Cycle Modeling}
\label{subsec:fine_grained_cost_model_numbers}
We instantiate the cost model of \autoref{sec:cost_model} with the following per-operation latencies, each determined by its implementation and by the routing path required.

\textbf{STAR (EFT).} To accurately model the injection failure in this protocol, we sample each injection's timing from the RUS process of~\cite{rescq}: preparation retries follow the post-selection and expansion success probabilities derived from ~\cite{rescq} as functions of $d$ and $p_{\text{ph}}$ ($d$=3 and $p_{ph}$ = $10^{-3}$), and teleportation succeeds with probability $1/2$, each failure requiring a fresh preparation at angle $2\theta$ until success or until the correction becomes Clifford. 

\textbf{Magic State Distillation (FFT).}
$T$ gates are realized via distillation and gate teleportation. We assume magic states are created quickly enough to be always available, so the modeled cost is the path-dependent routing of a state from its patch to the target qubit.

\begin{figure}[t]
    \centering
    \includegraphics[width=\columnwidth, clip]{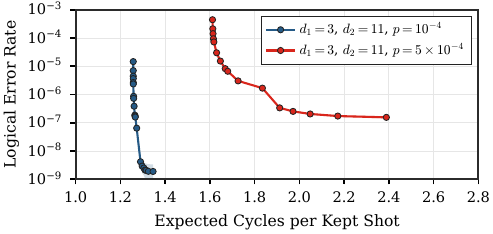}
    \caption{Expected cycles per kept $\ket{T}$ shot under MSC.}
    \label{fig:msc_sweep}
\end{figure}

\textbf{Magic State Cultivation (FFT).}
\label{para:magic_state_cultivation_numbers}
MSC operates in a RUS manner: each attempt takes $C_{\text{cult}}$ syndrome rounds and succeeds with acceptance probability $r_{\text{keep}}$. Here, $C_{\text{cult}}$ is fixed by the cultivation circuit, whereas $r_{\text{keep}}$ is determined by the target logical error rate (LER) of the output state, which we evaluate via end-to-end simulations. We select code distances $d_1 = 3$ and $d_2 = 11$, which yield the best latency--fidelity trade-off (\autoref{fig:msc_sweep}); this sets $C_{\text{cult}} = 13$, with $r_{\text{keep}} = 0.939$ for a target LER below $10^{-6}$ and $r_{\text{keep}} = 0.916$ for $10^{-8}$. We sample the latency of each $T$ gate (in logical cycles) as $\tau_T = N_{\text{att}} \cdot C_{\text{cult}}/d_2 + \tau_{\text{CNOT}} + B\,\tau_S$,
where $N_{\text{att}} \sim \mathrm{Geom}(r_{\text{keep}})$ is the number of attempts until acceptance, dividing by $d_2$ converts syndrome rounds to logical cycles, $\tau_{\text{CNOT}} = 2$~cycles (for a nearby patch) is the cost of performing a CNOT, and $B \sim \mathrm{Bernoulli}(1/2)$ indicates whether an $S$ correction ($\tau_S = 3$~cycles ~\cite{gidney2024inplace}) is required by the teleportation measurement. The latency of a synthesized $R_z(\theta)$ rotation sums $N_T$ independent draws of $\tau_T$ plus Clifford overhead. When C-Phases are applied in groups, a layer of concurrent rotations latency is determined by the slowest one. 


\textbf{$\boldsymbol{R_z(\theta)}$ Decomposition.}
For high-precision rotations we use GridSynth~\cite{gridsynth} to decompose $R_z(\theta)$ into $H$, $S$, and $T$ gates. We set the target synthesis error to be $p=10^{-3}$, with lower tolerance leading to more $T$ gates per rotation. $T$ gates use the LS protocols above, while $H$ and $S$ are performed~\cite{gidney2024inplace,litinski2019game}, at 1 and 3 clock cycles respectively~\cite{gidney2024inplace}. Our cost model can adapt to additional operations and alternative costs as QEC protocols improve.

 Our experiments were run on Perlmutter CPU nodes (2× AMD EPYC 7763, AVX2, 128 cores, 512 GB memory, 24-hour max runtime)~\cite{nersc_perlmutter}.
\section{Evaluation}
\subsection{Performance of Qomet in STAR}
\begin{figure}[t]
    \centering
    \includegraphics[width=\columnwidth,clip]{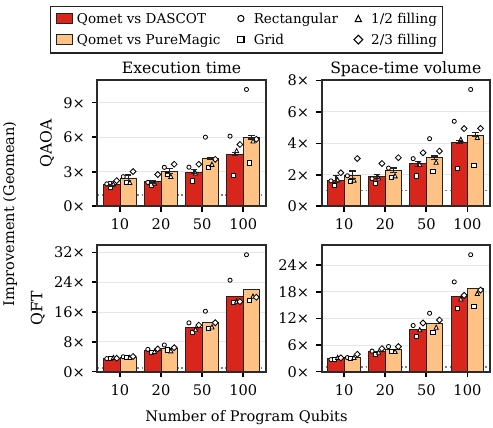}
    \caption{Performance improvements over the baselines for STAR across all layouts. Intervals represent the min-max.}
    \label{fig:Qomet_STAR}
\end{figure}

Figure~\ref{fig:Qomet_STAR} shows Qomet's reduction in execution time and space-time volume over DASCOT and PureMagic in the STAR architecture. It evaluates QAOA and QFT on all our four layouts. Qomet significantly outperforms the baselines for all these configurations. Across both programs and all four layouts, Qomet reduces execution time by up to $24.5\times$ over DASCOT and $31.4\times$ over PureMagic (geomean $4.99\times$ and $6.11\times$). We also reduce space-time volume by up to $20.2\times$ and $26.4\times$ (geomean $4.28\times$ and $4.89\times$). The gains grow with circuit size because Qomet has opportunities to merge C-Phase gates into one multi-target block. Its $R_z (\theta)$ gates then run in parallel, and each STAR injection is split across the free neighbouring patches. The baselines instead serialize two CNOTs and up to three $R_z (\theta)$ injections per gate, since gates within one C-Phase aren't allowed to commute much with gates from another C-Phase that share a control or target qubit. QFT benefits more than QAOA, because every qubit controls a longer chain of C-Phases in its circuit.  We also note that improvements in space-time volume are slightly less than execution time because Qomet's routes occupy more ancilla patches per operation. However, the volume is dominated by data qubits ($n \times T$), which shrinks with execution time, so the net reduction remains large.

\subsection{Performance of Qomet in the FFT regime}
\subsubsection{Magic State Cultivation}
\begin{figure}[t]
    \centering
    \includegraphics[width=\columnwidth,clip]{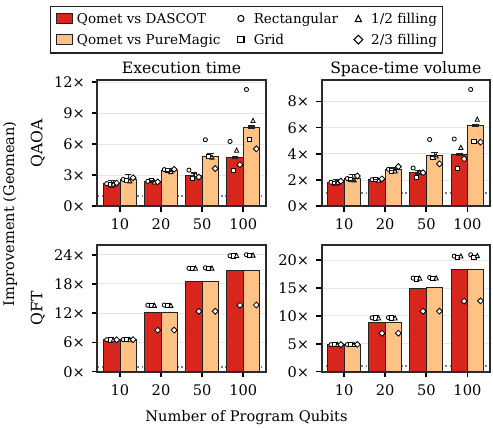}\
    \caption{Performance improvements over the baselines for MSC across all layouts. Intervals represent the min-max.}
    \label{fig:Qomet_MSC}
\end{figure}

Figure~\ref{fig:Qomet_MSC} shows Qomet's performance when magic states are produced by MSC. As in STAR, we evaluate QAOA and QFT on all four layouts. Qomet outperforms both baselines in every configuration, and its gains in MSC are the largest of the three regimes. Across both programs and all four layouts, Qomet reduces execution time by up to $23.9\times$ over DASCOT and $24.0\times$ over PureMagic (geomean $6.22\times$ and $7.51\times$). It also reduces space-time cost by up to $20.8\times$ and $21.0\times$ (geomean $5.08\times$ and $6.01\times$). In MSC every non-Clifford $R_z(\theta)$ is synthesized into several $T$ gates, and each $T$ state is cultivated on a neighbouring patch. An $R_z(\theta)$ is therefore far more expensive than CNOT, when compared to STAR, and the improvements are decided by how many $R_z(\theta)$ can run at once. We note that for QFT, DASCOT and PureMagic perform similarly at every size. Both execute each C-Phase as a fixed sequence of cultivated rotations along QFT's dependency chain, and better routing cannot shorten that cultivation latency, so both are limited by the same serialized cultivation. Qomet removes this bottleneck by reordering the commuting $CP(\theta)$.

\subsubsection{Magic State Distillation}

\begin{figure}[t]
    \centering
    \includegraphics[width=\columnwidth,clip]{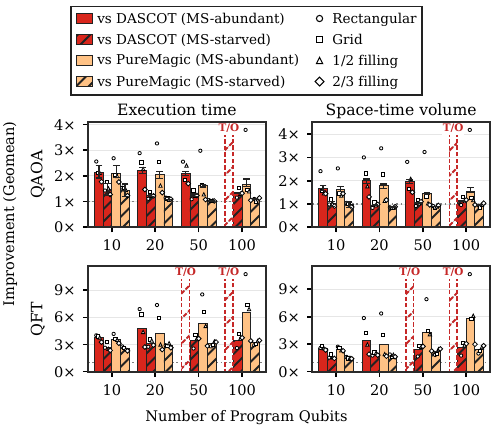}
    \caption{Performance improvements over the baselines for MSD ($\epsilon= 10^{-3}$). Intervals represent the min-max.}
    \label{fig:Qomet_MSD}
\end{figure}

Figure~\ref{fig:Qomet_MSD} illustrates results under MSD. This regime is harder for every compiler. Since each $\ket{T}$ state is
routed from the boundary and every $R_z(\theta)$ needs several such
routes, routing conflict governs execution time. With abundant supply, Qomet
reduces execution time by up to 6.9$\times$ over DASCOT and 10.8$\times$
over PureMagic (geomean 2.81$\times$ and 2.98$\times$), and space-time
volume by up to 5.9$\times$ and 10.7$\times$ (geomean 2.25$\times$ and
2.42$\times$). Gains are smaller than in STAR and MSC because Qomet's
parallel Stage~B sends all of a group's $|T\rangle$ requests to the
boundary at once, where they contend for the same factories and for cells
held by the CNOT stages' Steiner trees; the baselines spread their
requests over a longer schedule. With starved supply, reductions shrink to
up to 4.2$\times$ and 3.4$\times$ in time (geomean 2.04$\times$ and
1.83$\times$) and 3.2$\times$ and 2.8$\times$ in volume (geomean
1.47$\times$ and 1.32$\times$), as every compiler waits on an MS. What remains is the reduction in rotation count: for QFT,
Stage~D merges each control's corrections into one rotation, so Qomet
requests fewer $|T\rangle$ states and keeps its largest starved gains.
QAOA's RZZ terms need one rotation per gate in every compiler, making
Qomet weaker. With abundant supply on the 2/3-filling layout,
where each program qubit borders a single routing column and Steiner trees
block their own targets' $T$ routes, Qomet is only 1.05$\times$ faster
than PureMagic at 100 qubits and uses 0.96$\times$ the volume at 50
qubits. With starved supply, it is near equal in time
(1.00--1.09$\times$) and uses up to 17\% more volume at 20 qubits, as
Steiner trees hold cells while targets wait for $|T\rangle$ states.

\subsection{Performance on C-Phase-Sparse Benchmarks}
\label{sec:eval-cphase-sparse}

Algorithm-tailored compilation is important for practical quantum utility, much
as domain-specific accelerators are in classical computing. But such
specialization is only useful if it also provides benefits when a program lacks sequences of C-Phase gates. QAOA and QFT are nearly all C-Phase gates. Other important workloads mix a few
C-Phase blocks with long runs of other gates, or have none at all. We therefore
compile twelve such circuits (8--299 qubits) end to end with Qomet
and compare against DASCOT and PureMagic in (Table~\ref{tab:c-phase_sparse_benchmarks_table}). Qomet reduces execution time by up to $4.55\times$ (geomean $1.62\times$) over
DASCOT and up to $7.63\times$ (geomean $1.93\times$) over PureMagic, and
space-time volume by up to $4.05\times$ (geomean $1.62\times$) and up to
$5.56\times$ (geomean $1.92\times$), respectively across all our regimes. These
circuits have few long runs of consecutive C-Phase gates, so grouping alone has
little to merge. Most of the gain comes instead from Qomet's pipelined execution.
The pipelined scheduling in Qomet drops the slice-based abstraction and places every gate right after its predecessor is done, in program dependency order. This is why the four circuits with no
C-Phase gates still gain $1.06$--$1.99\times$. Our grouping adds to this where
C-Phase gates do cluster (sk-ising, draper, plane-wave), which gives the largest
gains. Qomet ties only on ham-sim-16-cp under MSC ($0.98\times$), and on
small CNOT-only circuits its volume is $0.83$--$0.98\times$ DASCOT's. DASCOT
times out under MSD on plane-wave-30 and draper-50, which Qomet compiles.
Qomet's benefits therefore extend beyond C-Phase-dense workloads. Our pipelined
execution speeds up general fault-tolerant circuits, and grouping adds further
gains wherever C-Phase gates cluster.

\begin{table*}[t]
    \centering
    \includegraphics[width=\textwidth,clip]{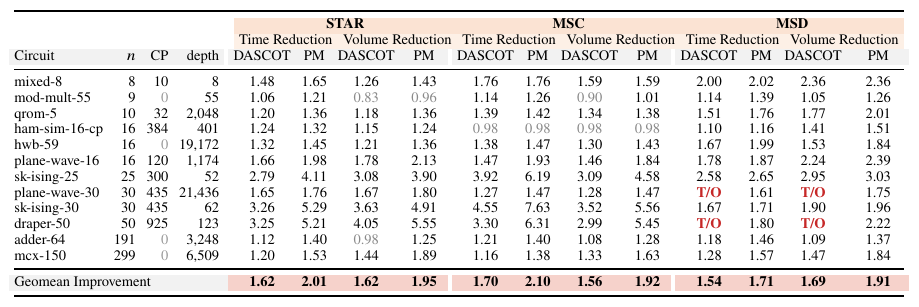}
    \caption{Performance of Qomet on C-Phase-Sparse Benchmarks against state-of-the-art FTQC Compilers. Each entry is the baseline's execution time (or space-time volume) divided by Qomet's, so values above $1.00$ favour Qomet.}
    \label{tab:c-phase_sparse_benchmarks_table}
\end{table*}

\subsection{PBC vs Clifford $+ T $ Execution Model Comparison}
\label{sec:eval-ppm-vs-ct}

\begin{figure}[t]
    \centering
    \includegraphics[width=\columnwidth,clip]{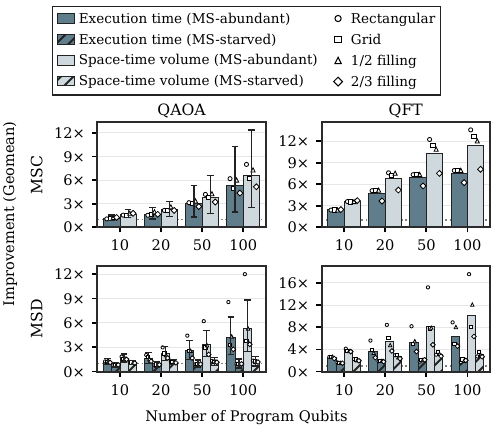}
    \caption{Comparison of execution methods for lattice surgery on the surface code.
    Clifford+T (Qomet) vs.\ Pauli-product measurements (PureMagic).}
    \label{fig:exec_method_ppm}
\end{figure}

So far we have compared Qomet only against other Clifford+T compilers. A
different execution model, Pauli-based computation (PBC), absorbs every Clifford
gate into the non-Clifford rotations and runs the program as a sequence of
multi-qubit PPMs. To compare the two execution
models end to end, we evaluate Qomet against PureMagic in \autoref{fig:exec_method_ppm} (with unrestricted Pauli weight), which is primarily designed for PBC. Both compilers run on the same layouts with the same
qubit placement and the same FFT magic-state model. Every route or PPM
costs two cycles in both compilers, so the comparison isolates the execution
model rather than any difference in routing cost \cite{litinski2019game,hofmeyr2025puremagic}. With abundant magic states, Qomet reduces execution time by up to $8.9\times$ (geomean $3.22\times$) and
space-time volume by up to $17.6\times$ (geomean $4.51\times$) across QAOA and
QFT, four layouts and 10--100 qubits. The gap widens with circuit size, from
$1.2$--$2.6\times$ at 10 qubits to $4.3$--$7.5\times$ at 100 qubits. Absorbing
the Cliffords removes them from the schedule, but the remaining rotations become
high-weight Paulis that touch most of the qubits. Each PPM then spans a large
part of the layout and blocks the others, so PBC executes the program almost
serially. Qomet keeps every operation local: a C-Phase group touches only its
control and targets, so independent groups and their rotations run in parallel.
With starved MSD, both models wait on the same four factories, so this
parallelism matters less: Qomet still reduces execution time by up to
$2.2\times$ (geomean $1.27\times$) and space-time volume by up to $3.6\times$
(geomean $1.79\times$).

On some QAOA instances PBC is competitive (0.80--0.95$\times$ in time,
1.05--1.35$\times$ in volume), depending on the synthesized $R_z(\theta)$.
GridSynth writes $R_z(\theta)$ as $\pi/8$ rotations followed by a
$\theta$-dependent Clifford $C$, which PBC must commute past the RZZ's
second CNOT. If $C$ commutes with it (a Pauli or $\sqrt{X}$), the CNOT
pair cancels, every rotation stays weight 2, and PBC wins. Otherwise the
CNOT pair survives as entangling Cliffords that widen later rotations up to
weight $\sim$55; these share qubits and run almost serially, so Qomet wins
by up to 1.5$\times$ despite equal $T$ counts. This affects almost 40\% of
instances, explaining the wide spread. PBC is thus competitive only when
magic-state supply is the bottleneck and the synthesized rotations keep
its Pauli weights low.

\subsection{Reduction in Program Failure Rate}
\begin{figure}[t]
    \centering
    \includegraphics[width=\columnwidth,clip]{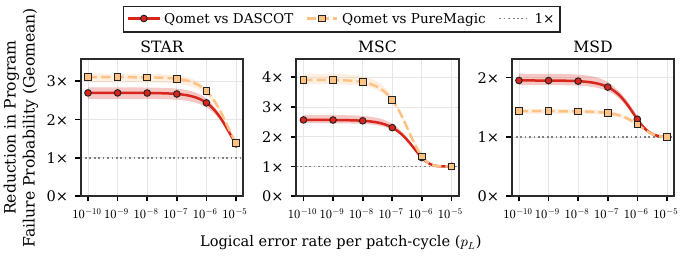}
    \caption{Reduction in program failure probability with Qomet over the baselines for
    50-Qubit QAOA as $p_L$ varies. Shaded bands represent min--max.}
    \label{fig:Program_Failure_Rate}
\end{figure}

~\autoref{fig:Program_Failure_Rate} shows $P_{\text{rel}}$ for 50-qubit QAOA across four layouts
as $p_L$ varies. In the low-$p_L$ regime that fault-tolerant hardware targets
($p_L = 10^{-9}$), Qomet reduces the program's failure probability by up to
$3.91\times$ (geomean $2.59\times$) over PureMagic and up to $2.70\times$
(geomean $2.38\times$) over DASCOT. At small $p_L$,
$P_{\text{fail}} \approx p_L V$, so $P_{\text{rel}}$ tracks Qomet's space-time
volume improvements closely. As $p_L$ grows toward $10^{-5}$, both compilers'
programs almost certainly fail, so $P_{\text{rel}}$ falls toward 1. Every cycle
and patch Qomet saves is one fewer opportunity for a logical fault, so it also
raises the chance that a single run succeeds. At low $p_L$, Qomet's volume reduction over the baselines improves reliability as much as building hardware with a proportionally lower
logical error rate, without the larger code distance and extra physical qubits
that would normally require.

\subsection{Compilation Time Comparison}
\label{sec:eval-compile-time}
\begin{figure}[t]
    \centering
    \includegraphics[width=\columnwidth,clip]{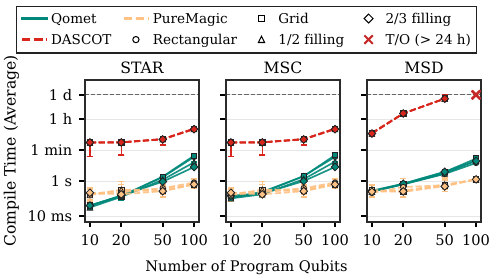}
    \caption{Compile-time scaling of Qomet and the baselines with the number of program qubits (QAOA).}
    \label{fig:compilation_time}
\end{figure}

Compilation can be a classical cost, but a fault-tolerant compiler must
still scale to the large programs FTQC targets. Fig.~\ref{fig:compilation_time}
reports compilation time for QAOA from 10 to 100 qubits on all four layouts. Qomet's compilation time grows polynomially with program size and stays under
32\,s at 100 qubits in every regime. Its work is one Steiner tree per C-Phase
group plus one greedy pass that places each group on the timeline, so it grows
with the number of groups and the grid area. DASCOT has a large fixed cost
(about 3 minutes even at 10 qubits), and scales poorly due to its optimizer-based search. Under STAR and MSC, its time stays flat up
to 50 qubits, then rises to about 16 minutes at 100 qubits. Under MSD,
where its search must also route every T state, its time grows
by roughly $10\times$ with each step in size, reaching about 15 hours at 50
qubits and timing out at 100 qubits. PureMagic grows the most slowly
(about 1\,s at 100 qubits). Qomet's compilation cost therefore grows smoothly with program size and remains practical at the scale of fault-tolerant programs.

\subsection{Effect of $R_z$-Aware Grouping}
To isolate grouping, we replace it with a naive greedy grouping that
uses C-Phase commutativity, symmetry, and fan-out
CNOTs~\cite{jin_structured_2022,herr2017lattice} but ignores layout and
routing cost (each group starts from the first ungrouped gate in program
order), leaving the rest of the pipeline unchanged. Across 20 random
30-qubit QAOA instances per configuration (Table~\ref{tab:grouping_ablation_30q}),
$R_z$-aware grouping is faster in 286 of 320 runs and 15 of 16
configurations, reducing execution time by 13.6\% (geomean). Gains are
largest under STAR and MSC (17.9\% and 21.3\%), where Stage~B dominates
latency, and about 7\% under MSD, where $|T\rangle$ routing bounds
latency. The only slowdown (1.2\%, MSD-abundant, 1/2-filling) is a
near-tie where a large group's targets compete for few free ancillas and
tight factory routes.

\begin{table}[t]
\caption{Execution-time reduction (\%) of $R_z$-aware over naive greedy
grouping on 30-qubit QAOA.
Negative means greedy does better.}
\label{tab:grouping_ablation_30q}
\centering
\footnotesize
\setlength{\tabcolsep}{5pt}
\begin{tabular}{@{}l rrrr r@{}}
\toprule
\textbf{Regime} & Rect. & Grid & 1/2 fill & 2/3 fill & \textbf{All} \\
\midrule
STAR           & 22.6 & 21.6 & 18.4   &  8.1 & \textbf{17.9} \\
MSC            & 23.3 & 28.9 & 22.0   &  9.8 & \textbf{21.3} \\
MSD (abundant) & 16.2 &  6.6 & $-$1.2 &  5.2 & \textbf{6.9}  \\
MSD (starved)  & 14.8 &  7.8 &  3.7   &  2.2 & \textbf{7.3}  \\
\bottomrule
\end{tabular}
\end{table}
\section{Discussion and Related Work}
\label{sec:discussion}
Prior FT compilers either use slice-based schedules~\cite{dascot}, abstract executable resource allocation~\cite{rescq}, or target EFT-specific resource management~\cite{eft_vqa}. Qomet instead combines a layout-aware cost model, LS-native multi-target interactions, and cycle-accurate scheduling across FTQC. Although Qomet targets C-Phase gates, the opportunity is organized by diagonality rather than by any single gate. Operations diagonal in the computational basis commute regardless of support overlap, so an entire diagonal block collapses to one \emph{phase polynomial}~\cite{Amy_2018} that can be resynthesized as a unit~\cite{van2020circuit}. This covers $R_z$ and $R_{zz}$ rotations, the $Z$-string rotations recurring in Ising cost layers and Trotterized simulation, and the multi-controlled phases ($C^kZ$) that dominate IQP sampling circuits~\cite{IQP_FT}; mixed-Pauli Hamiltonians retain partial freedom through partitioning into commuting subsets. Gate-level decompositions hide this freedom behind overlapping CNOT ladders, and the payoff under lattice surgery is space-time volume rather than gate count. Qomet's scheduler also maps naturally onto dynamic and real-time execution: the
feed-forward it already handles (Pauli-frame updates, conditional $S$
corrections, RUS retries) is sampled offline here, but groups and routes could
be precomputed while dispatch reacts to measured outcomes at runtime.
\section{Conclusion}
Realizing FTQC requires efficient logical compilation. 
However, current slice-based compilers ignore hardware latencies, causing severe resource underutilization. 
We introduce \textbf{Qomet}, a microarchitecture-aware compiler that exploits algorithmic structure for multi-target interactions and uses cycle-accurate scheduling to eliminate false dependencies. This unlocks parallelism, achieving significant speedups over SoTA and scaling to complex workloads.
\section{Acknowledgments}

This material is based upon work supported by the U.S. Department of Energy, Office of Science, Office of Advanced Scientific Computing Research, Accelerated Research in Quantum Computing under Award Number DE-SC0025633.
This research was, in part, supported by the DARPA HARQ Program, under Agreement No. HR0011-26-9-E096.
This material is based upon work supported by the U.S. National Science Foundation under Cooperative Agreement No. 2547483 through a subaward to the University of Michigan.
This research used resources of the National Energy Research Scientific Computing Center, a DOE Office of Science User Facility supported by the Office of Science of the U.S. Department of Energy under Contract No. DE-AC02-05CH11231 using NERSC award NERSC DDR-ERCAP0035341.
This research was, in part, funded by the U.S. Government. The views and conclusions contained in this document are those of the authors and should not be interpreted as representing the official policies, either expressed or implied, of the U.S. Government. LLMs were used in this project to assist with organizing the codebase and with scripting and managing experiment runs on HPC systems. All research ideas, methods, analyses, and conclusions are the authors’ own.

\bibliographystyle{ACM-Reference-Format}
\bibliography{refs_dhanvi}

\end{document}